\documentclass[sigconf]{acmart}

\AtBeginDocument{%
  \providecommand\BibTeX{{%
    \normalfont B\kern-0.5em{\scshape i\kern-0.25em b}\kern-0.8em\TeX}}}

\setcopyright{acmcopyright}
\copyrightyear{2025 Copyright is held by the owner/author(s). Publication rights licensed to}
\acmYear{2025}
\acmDOI{10.1145/3728425.3759905}

\acmConference[SVC '25]{the 1st Workshop \& Challenge on Subtle Visual Computing, 
co-located with the 33rd ACM International Conference on Multimedia}
{October 27--31, 2025}{Dublin, Ireland}
\acmPrice{15.00}
\acmISBN{979-8-4007-1837-3/2025/10}

\usepackage{graphicx}
\usepackage{multirow}
\usepackage{enumitem}
\usepackage{bbding}
\usepackage[capitalize]{cleveref}
\crefname{section}{Sec.}{Secs.}
\Crefname{section}{Section}{Sections}
\Crefname{table}{Table}{Tables}
\crefname{table}{Tab.}{Tabs.}
 
\usepackage{amsthm,amsmath,amssymb}
\usepackage{mathrsfs}
\usepackage{booktabs}
\usepackage{multicol}
\usepackage{color}
\usepackage{hyperref}
\usepackage{float}
\usepackage{subfig}
\usepackage{rotating}
\usepackage{booktabs}   % 提供 \toprule, \midrule, \bottomrule 等三线表命令
\usepackage{multirow}   % 支持表格中单元格的纵向合并
\usepackage{graphicx}   % 如果表格中使用了 \resizebox（可选）
\usepackage{colortbl}   % 如果你想加颜色（例如高亮最优结果，可选）
\usepackage{arydshln}
\usepackage{pifont} % for checkmarks and crosses

\definecolor{LightCyan}{rgb}{0.88,1,1}

\begin{document}

\title{AD-AVSR: Asymmetric Dual-stream Enhancement for Robust Audio-Visual Speech Recognition}

\author{Junxiao Xue}
\affiliation{%
  \institution{Zhengzhou University}
  \city{Zhengzhou}
  \country{China}
}
\affiliation{%
  \institution{Zhejiang Lab}
  \city{Hangzhou}
  \country{China}
}
\email{xuejx@zhejianglab.cn}
\authornote{Both authors contributed equally to this research.}

\author{Xiaozhen Liu}
\affiliation{%
  \institution{Zhengzhou University}
  \city{Zhengzhou}
  \country{China}
}
\email{liuxiaozhen123@gs.zzu.edu.cn}
\authornotemark[1]
\authornote{Corresponding author.}

\author{Xuecheng Wu}
\affiliation{%
  \institution{Xi'an Jiaotong University}
  \city{Xi'an}
  \country{China}
}
\email{xuecwu@gmail.com}

\author{Xinyi Yin}
\affiliation{%
  \institution{Zhengzhou University}
  \city{Zhengzhou}
  \country{China}
}
\email{yinxinyi@stu.zzu.edu.cn}

\author{Danlei Huang}
\affiliation{%
  \institution{Xi'an Jiaotong University}
  \city{Xi'an}
  \country{China}
}
\email{forsummer@stu.xjtu.edu.cn}

\author{Fei Yu}
\affiliation{%
  \institution{Zhejiang Lab}
  \city{Hangzhou}
  \country{China}
}
\email{yufei@zhejianglab.com}

% \affiliation{%

% \affiliation{%

% \affiliation{%

% \affiliation{%

% \affiliation{%

% \affiliation{%

\begin{abstract}
Audio-visual speech recognition (AVSR) combines audio-visual modalities to improve speech recognition, especially in noisy environments. However, most existing methods deploy the unidirectional enhancement or symmetric fusion manner, which limits their capability to capture heterogeneous and complementary correlations of audio-visual data—especially under asymmetric information conditions. To tackle these gaps, we introduce a new AVSR framework termed AD-AVSR based on bidirectional modality enhancement. Specifically, we first introduce the audio dual-stream encoding strategy to enrich audio representations from multiple perspectives and intentionally establish asymmetry to support subsequent cross-modal interactions. The enhancement process involves two key components, \textit{i.e.}, Audio-aware Visual Refinement Module for enhanced visual representations under audio guidance, and Cross-modal Noise Suppression Masking Module which refines audio representations using visual cues, collaboratively leading to the closed-loop and bidirectional information flow. To further enhance correlation robustness, we adopt a threshold-based selection mechanism to filter out irrelevant or weakly correlated audio-visual pairs. Extensive experimental results on the LRS2 and LRS3 datasets indicate that our AD-AVSR consistently surpasses SOTA methods in both performance and noise robustness, highlighting the effectiveness of our model design.
\end{abstract}


% ,Multimedia information systems


\maketitle

%\renewcommand{\thefootnote}{\fnsymbol{footnote}}
%\footnotetext[1]{indicates corresponding authors.}

\section{Introduction}
\label{sec:introduction}
Audio-Visual Speech Recognition (AVSR), which recovers the natural language expressions from the audio-visual inputs, offers broad applications \cite{sun2018lip, burchi2023audio, ryumin2024audio}. By leveraging more information than video-only or audio-only methods \cite{martinez2020lipreading, kim2022distinguishing}, AVSR delivers superior performance. However, naively fusing information from different modalities and constructing a mapping from the combined features to natural language often fails to achieve the desired "1+1>2" effect. This arises from two main reasons: First of all, the information from different modalities is not symmetrical; a single frame of visual information may correspond to multiple frames of audio information, and vice versa. This asymmetry makes it challenging for existing strategies like unidirectional enhancement \cite{hong2022visual, xu2020discriminative, kim2024learning} or visually driven speech reconstruction \cite{yang2022audio, hsu2023revise} to robustly establish inter-modal connections. Second, various real-world application scenarios suffer from audio-visual asynchrony due to recording equipment and environmental constraints, resulting in extensive irrelevant audio-visual pairs. Existing AVSR methods \cite{ma2021end, hong2023watch, hu2023cross} do not effectively filter out the irrelevant pairs.

% \Rz{To address the above challenges, we introduce a dual-stream network designed for directional multi-modal asymmetric fusion. Our key idea involves constructing asymmetric bidirectional enhancement: using excessive audio when audio enhances video and excessive video when video enhances audio, enabling the model to better understand the relationships between modalities.}
To address the above challenges, we introduce a dual-stream network designed for directional multi-modal asymmetric fusion. Our key idea involves constructing asymmetric bidirectional enhancement and fusion, which primarily consists of three stages.
In the first stage, to enable effective bidirectional cross-modal enhancement, we introduce an audio dual-stream encoding strategy during initial preprocessing and feature extraction. This strategy uses excessive audio when audio enhances video and excessive video when video enhances audio, enabling the model to better understand the relationships between modalities. The second stage is the bidirectional enhancement module, where we introduce two synergistic submodules. The first, the Audio-aware Visual Refinement Module (AVRM), divides each video frame into multiple visual regions, adaptively adjusting the weights of these regions based on audio occurrences to accurately capture lip movement representations. The second submodule is the Cross-modal Noise Suppression Masking Module (CMNSM), which first merges visual and audio information using traditional cross-modal attention. It then leverages visual information to suppress redundant audio noise and assess the reliability scores at each time step, forming a bidirectional interactive loop with the first module.

Additionally, to address the interference caused by many irrelevant audio-visual pairs due to asynchrony between audio and video, the third stage introduces a threshold-based selection mechanism (TBSM) \cite{zhou2021positive} to ensure that only positive audio-visual connections are retained during fusion. This mechanism allows the network to utilize the similarities between audio-visual pairs to learn more representative features, filtering out negative and weaker similarities for more efficient fusion. 

Beyond the task of speech recognition, subtle lip movements also convey fine-grained visual cues such as micro-expressions. Although weak visual signals may have limited salience, they can significantly enhance the model’s understanding of user intent in noisy or semantically ambiguous scenarios, making them valuable for applications like deception detection \cite{guo2023audio, guo2024benchmarking} and robust multimodal learning~\cite{hong2023watch}. Our contributions can be summarized in four key aspects: 
\begin{itemize}[leftmargin=*]
\item We propose AD-AVSR, a method that leverages the complementary strengths of visual and audio modalities to enhance cross-modal fusion. It introduces an innovative audio dual-stream encoding strategy that enhances audio representations via joint time-frequency modeling, while explicitly establishing asymmetric interactions between modalities.
\item We propose a Bidirectional Modality Enhancement module to improve AVSR fusion, consisting of two parallel cross-modal components: the Audio-aware Visual Refinement module and the Cross-modal Noise Suppression Masking module.
\item We mitigate interference from irrelevant audio-visual pairs by introducing a threshold-based selection mechanism that improves fusion efficiency through pairwise similarity evaluation.
    % \item We introduce two innovative submodules—the Audio-aware Visual Refinement module and the Cross-modal Noise Suppression Masking module. The former adaptively adjusts visual weights using audio information, while the latter utilizes visual information to suppress noise in audio interference. 
    % \item Additionally, to alleviate the interference caused by irrelevant audio-visual pairs during the fusion process, we also introduce a threshold-based selection mechanism that optimizes feature fusion based on the similarity of audio-visual pairs, effectively enhancing the fusion efficiency of the AVSR system.
\item Our proposed AD-AVSR achieves state-of-the-art results on the LRS2 and LRS3 datasets, demonstrating its outstanding efficacy in the AVSR task.
\end{itemize}

\section{Related Works}
\label{sec:related}
\subsection{Audio Visual Speech Recognition} \label{subsec:avsr}
With the rapid development of deep learning, Automatic Speech Recognition (ASR) systems have made great strides and found widespread use in real-world applications \cite{ivanko2023review, karbasi2022asr, bhardwaj2022automatic, ma2024characterization}. However, relying solely on audio can be limiting, especially when the signal is noisy or missing \cite{sheng2024deep}. To address this, Audio-Visual Speech Recognition (AVSR), which integrates both audio and visual inputs, has emerged as a key research focus \cite{ma2021end, hong2022visual, hong2023watch, yu2024hourglass, wang2025dcim}. Early AVSR studies \cite{afouras2018deep} typically extracted features from both modalities, fused them at the feature level, and then applied sequence modeling to the combined representation. For example, Chung et al. \cite{son2017lip} proposed a "Watch, Listen, Attend and Spell" (WLAS) network and introduced a large-scale lip-reading sentence dataset, LRS2. Ma et al. \cite{ma2021end} proposed replacing traditional recurrent networks with the Conformer architecture and combining CTC and attention loss, which significantly improved performance. Shi et al. \cite{shi2022robust} proposed AV-HuBERT, a self-supervised framework that exploits the strong correlation between audio and lip movements for audio-visual representation learning. 
Burchi et al. \cite{burchi2023audio} recently proposed using patch-wise attention instead of component-wise attention, achieving similar performance with reduced complexity. 
In recent years, Zhang et al. \cite{zhang2024visual} explored AVSR inference under fully missing visual input and introduced a visual generation model based on discrete features. In the same year, Wang et al. \cite{wang2024restoring} proposed AVLR, an audio-visual lip restoration framework that mitigates the impact of lip occlusion by reconstructing the masked mouth regions.

Subsequently, many studies have explored enhancement strategies before audio-visual feature fusion, which can be grouped into three types. The first uses visual information to enhance audio features—for example, V-CAFE \cite{hong2022visual} improves robustness by injecting visual cues into audio, and Bo Xu et al. \cite{xu2020discriminative} separated speech from noise using lip movements before fusion. The second leverages audio cues to improve visual modeling, such as the cross-modal attention mechanism in \cite{kim2024learning}. The third focuses on visual-driven speech modeling, including neural vocoder-based reconstruction \cite{yang2022audio} and the ReVISE model \cite{hsu2023revise}, which enables speech resynthesis from visual input. While effective, these methods still face challenges in fully capturing the heterogeneity and complementarity of modalities, especially under asymmetric or noisy conditions. To tackle this, we propose a bidirectional modality enhancement framework—AD-AVSR—which jointly optimizes structural and semantic fusion for improved multimodal integration.
\subsection{Cross-modal Enhancement and Fusion} \label{subsec:cmef}
To overcome the limitations of existing AVSR systems, recent studies have focused on cross-modal enhancement mechanisms. For example, Hu et al. \cite{hu2023cross} proposed the GILA framework, which models modality complementarity through global interaction and captures frame-level temporal consistency via local alignment. Other works, such as \cite{zhang2022learning}, employ Transformer-based cross-modal attention with adaptive masking to achieve deep feature fusion. MLCA-AVSR \cite{wang2024mlca} enables each modality to learn complementary context from the other across feature hierarchies, enhancing representation learning. CATNet \cite{wang2024catnet} introduces a dual-modality network to extract key features and suppress noise, improving collaboration between modalities. Hong et al. \cite{hong2023watch} proposed AV-RelScore, an audio-visual reliability scoring module that evaluates the reliability of each input modality prior to fusion. This enables robust speech recognition even when one or both modalities are degraded. Fu et al. \cite{fu2024boosting} proposed PCD, the first work dedicated to enhancing the robustness of AVSR under dual-modality distortion. 

Related fields have also explored cross-modal enhancement and fusion. Yang et al. \cite{yang2024cmaf} proposed a modality enhancement method that exploits the complementarity between enhanced sonar features and raw visual features to compensate for the limited image information in acoustic-only scenarios. Pang et al. \cite{pang2022heterogeneous} introduced Weak Semantic Augmentation (WSA), a novel text enhancement module for image retrieval that enriches textual features using relevant visual information to build a high-level semantic space.  Inspired by these approaches, we propose two cooperative modules: an Audio-aware Visual Refinement Module and a Cross-modal Noise Suppression Masking Module, which enhance cross-modal modeling and information selection in AVSR systems.

\section{Methodology}
\label{sec:method}
\begin{figure*}[th]
    \centering
    \includegraphics[width=0.85\linewidth]{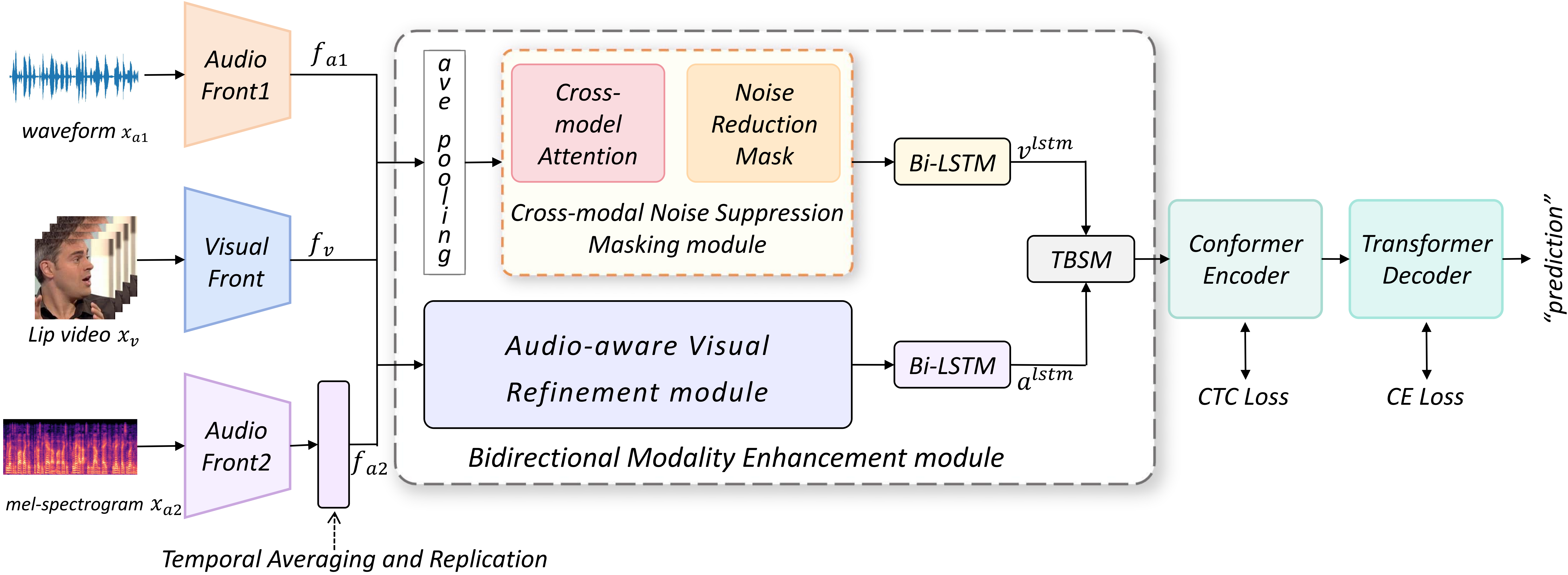} % 替换为你的图片文件名
    \caption{Overall architecture of the proposed AD-AVSR. TBSM refers to the Threshold-based Selection Mechanism.}
    \label{fig:framework}
\end{figure*}

Let the preprocessed input data consist of a lip-reading video sequence $x_v \in \mathbb{R}^{T_0 \times H_0 \times W_0 \times C_0}$,  the waveform representation $x_{a1} \in \mathbb{R}^{S'}$, the log-Mel spectrogram derived from the corresponding audio signal $x_{a2} \in \mathbb{R}^{F \times S}$, and the corresponding ground-truth transcription $y \in \mathbb{R}^{L}$. Here, $T_0$ denotes the total number of video frames, and $H_0$, $W_0$, and $C_0$ represent the height, width, and number of channels of each frame, respectively. $F$ and $S$ refer to the number of frequency bins and time steps in the spectrogram, while $S'$ denotes the length of the raw waveform. $L$ corresponds to the length of the transcribed sentence. 
To enable effective bidirectional modality enhancement, we adopt an audio dual-stream encoding strategy that captures both spectral and temporal cues, enriching audio representations and introducing intentional modality asymmetry. The encoded audio and visual features are fed into a bidirectional cross-modal enhancement module. In the visual-enhanced audio branch, the Cross-modal Noise Suppression Masking Module (CMNSM) uses audio queries to attend to informative visual features with a gating mechanism to suppress noise. In the audio-enhanced visual branch, the Audio-aware Visual Refinement Module (AVRM) partitions frames into k regions and uses visual queries to focus on articulation-relevant areas, alleviating occlusion and sync issues with low parameter cost. A Threshold-based Selection Mechanism (TBSM) then prunes weakly correlated audio-visual pairs, ensuring strong cross-modal consistency. These modules—CMNSM for audio-guided fusion, AVRM for visual grounding, and TBSM for consistency refinement—produce multimodal features for sequence modeling, followed by prediction via a Transformer decoder.
% To better support the subsequent bidirectional modality enhancement modules, we introduce an audio dual-stream encoding strategy, which models the audio signal from both spectral and temporal perspectives. This enriches the audio representation and introduces intentional modality asymmetry. The encoded audio and visual features are then passed into our bidirectional cross-modal enhancement module, where cross-modal interactions between the two modalities are performed to facilitate mutual reinforcement. Specifically, Cross-modal Noise Suppression Masking
% Module (CMNSM, visual-enhanced audio branch) uses audio queries to attend to informative visual features and includes a gating mechanism to suppress irrelevant noise. Audio-aware Visual Refinement Module (AVRM, audio-enhanced visual branch) divides visual frames into k regions and uses visual queries to locate articulation-relevant areas, mitigating occlusion and sync issues with low parameter cost.
% Following this, the subsequent Threshold-based Selection Mechanism (TBSM) further prunes the connections, preserving only strongly correlated audio-visual pairs.
% % Following this, we introduce a threshold-based selection mechanism (TBSM) to filter out irrelevant or weakly correlated audio-visual pairs.
% These modules are complementary: CMNSM supports audio-guided fusion, AVRM aids visual grounding, and TBSM refines cross-modal consistency. The resulting multimodal features are subsequently fed into a sequence modeling module, where loss is computed. Finally, the prediction is generated by a Transformer decoder. 
The overall AD-AVSR framework is shown in Figure \ref{fig:framework}.
\subsection{Audio Dual-Stream Encoding} \label{subsec:dual-encoding}
In traditional audio-visual speech recognition (AVSR) tasks, audio information is typically modeled using a single-stream encoding strategy. However, this method is not well-suited for our proposed bidirectional modality enhancement framework, primarily for the following reasons.
The typical frame rate for video is 25 fps, while the audio sampling rate is 16,000 Hz. This implies that in the raw input, each video frame inherently contains more information than the corresponding audio segment. Despite temporal dimension compression through convolutional feature extraction later aligning audio with visual modalities in terms of frame length and dimensional features, this alignment does not meet our structural needs for information density differences during the enhancement phase. Our objective is to maintain a lower information density in the modality being enhanced and a higher density in the enhancing modality during cross-modal interactions, facilitating effective supplemental guidance.
Therefore, we employ two audio encoding strategies originating from the same source. 

The first method utilizes time-domain encoding, which converts the original audio into a waveform followed by feature extraction via a 1D CNN combined with ResNet18, as referenced in \cite{hong2023watch}. This process results in audio features denoted as \( f_{a1} \in  \mathbb{R}^{T_1 \times C_1} \), and visual features \( f_v \in \mathbb{R}^{T_1 \times C_1 \times H_1 \times W_1} \). This encoding strategy is primarily designed to use visual information to guide audio enhancement, facilitating the capture of more granular dynamic changes in each visual frame, thereby compensating for the audio’s lack of short-term detail accuracy. The second encoding strategy is frequency-domain encoding, intended to enhance visual information with audio features. In this approach, the original audio is first converted into a waveform, then transformed into a Mel spectrogram via Short-Time Fourier Transform (STFT). After straightforward processing through a 1D convolution, the dimensions of the audio output are adjusted to align with the visual information. Furthermore, to ensure that each frame's audio information contains more visual data, every adjacent set of 25 audio frames is summed, averaged, and this process is repeated 25 times to enhance the visual relevance of the audio. 
The final output audio features \( f_{a2} \) also exhibit the shape \( \mathbb{R}^{T1 \times C1} \).

\subsection{Cross-modal Noise Suppression Masking Module } \label{subsec:video-to-audio}
\begin{figure}[t!]
    \centering
    \includegraphics[width=0.8\linewidth]{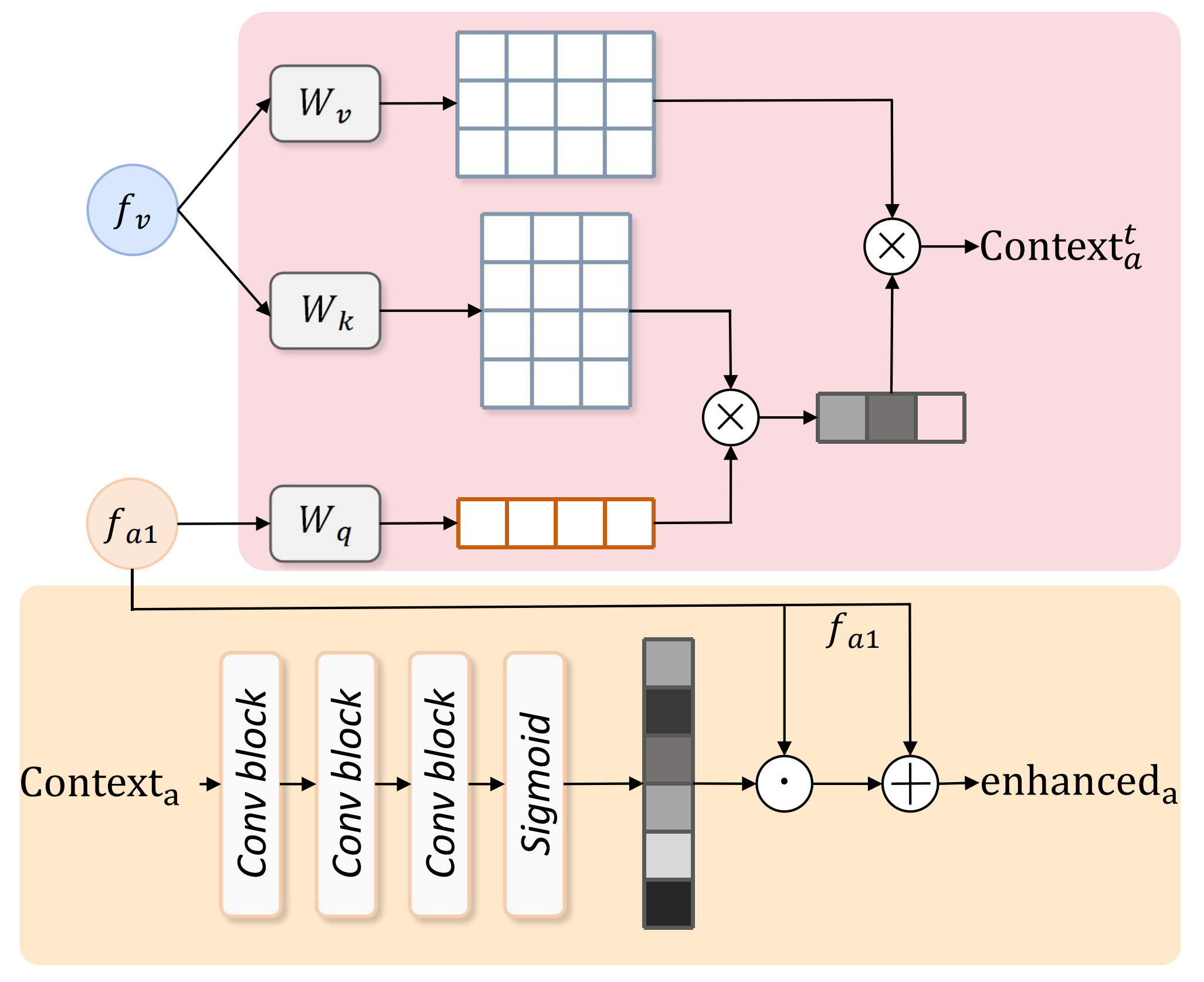} % 替换为你的图片文件名
    \caption{Detailed architecture of cross-modal noise suppression masking
module.}
    \label{fig:CMNSM}
\end{figure}

Audio signal recognition is not only influenced by its positional information but also significantly by the linguistic context, which is a critical factor in speech recognition tasks. Therefore, in audio-visual speech recognition (AVSR) tasks, understanding the movements of the lips before and after a given moment is beneficial for enhancing model performance. Additionally, audio information is often disrupted by environmental noise, which severely impacts the accuracy and robustness of recognition. To address these challenges, we propose a Cross-Modal Noise Suppression Masking Module (CMNSM). In this module, we employ a cross-modal attention mechanism to integrate both local and global movements of the lips along with visual background information to enrich the audio signal representation. The masking suppression module further mitigates the adverse effects of environmental noise on the audio signals. As illustrated in Figure \ref{fig:CMNSM}, the features of each modality---audio features \( f_{a1} \in \mathbb{R}^{T_1 \times C_1} \) and visual features \( f_v \in \mathbb{R}^{T_1 \times C_1 \times H_1 \times W_1} \)---are obtained through specific front-end embedding techniques. Subsequently, our attention module captures the visual context at each time step \( t \) through cross-modal attention, as described below:
\begin{equation}
\text{Context}_a = \text{softmax}\left(\frac{Q^t \cdot K^T}{\sqrt{D_1}}\right) \cdot V,
\end{equation}
\begin{equation}
Q^t = f_{a1}^t \cdot W_q, \quad K = f_v \cdot W_k, \quad V = f_v \cdot W_v,
\end{equation}
\( \text{Context}_{a}^{t} \) represents the context representation of audio features at time step $t$, extracted based on visual information. $Q$, $K$, and $V$ denote the Query, Key, and Value matrices in cross-modal attention, respectively. The embedding weights for query, key, and value are $W_q \in \mathbb{R}^{C_1 \times D_1}$, $W_k \in \mathbb{R}^{C_1 \times D_1}$, and $W_v \in \mathbb{R}^{C_1 \times D_1}$. The Query matrices is derived from audio information, while the Key and Value matrices are obtained from visual information. After cross-modal attention, the audio features contextualized by visual information are represented as $\text{Context}_a = \{\text{Context}_a^{1}, \text{Context}_a^{2}, \ldots, \text{Context}_a^{T_1}\} \in \mathbb{R}^{T_1 \times D_1}$.

The audio $\text{Context}_a$ is then processed further in the Noise Suppression Masking Module, which incorporates a learnable mask generator with two primary objectives: firstly, to suppress noise components within the audio features through predicted masks, enhancing the system's robustness; secondly, the mask generator performs temporal convolution modeling across the time dimension of the audio sequence. The generated masks also serve as confidence scores for each timestep, ranging from 0 to 1---where 0 indicates low reliability of the speech modeling at that timestep, and 1 indicates complete reliability. This mask generator consists of three convolutional layers, each employing ReLU and Softmax activation functions. The detailed computational process is described as follows:
\begin{align}
m_0 &= \text{ReLU}\Big( \text{BN}\Big( \text{Conv}(\text{Context}_{a}) \Big) \Big), \\
% m_1 = \text{ReLU}\Big( \text{BN}\Big( \text{Conv}(m_0) \Big) \Big), 
% m = \text{Sigmoid}\Big( \text{BN}\Big( \text{Conv}(m_1) \Big) \Big).
m &= \text{Sigmoid}\Big( \text{BN}\Big( \text{Conv}\Big( \text{ReLU}\Big( \text{BN}\Big( \text{Conv}(m_0) \Big) \Big) \Big) \Big) \Big).
% m &= \text{Sigmoid}\Big( \text{BN}\Big( \text{Conv}\Big( \text{ReLU}\Big( \text{BN}\Big( \text{Conv}\Big( \nonumber \\
% &\quad \text{ReLU}\Big( \text{BN}\Big( \text{Conv}(\text{Context}_{a}) \Big) \Big) \Big) \Big) \Big) \Big) \Big) \Big).
\end{align}

Here, BN stands for Batch Normalization, a process used to standardize the inputs to a network, promoting faster and more stable training. The resulting confidence score mask \( m \) also has a tensor dimension of \( T_1 \times D_1 \). In the subsequent experiments, the dimensions of \( C_1 \) and \( D_1 \) are kept consistent to ensure uniformity and effectiveness in feature processing.

The generated mask \( m \) is multiplied with the input audio features \( f_{a1} \). With the aid of visual information, the noise components in the audio are effectively suppressed, while the crucial speech information remains largely unchanged. Subsequently, this processed audio information is combined with the original audio features \( f_{a1} \), resulting in the enhanced audio output \( \text{enhanced}_{a} \in \mathbb{R}^{T_1 \times D_1} \)  . The operational formula is illustrated in the following diagram:
\begin{equation}
\text{enhanced}_{a}  = (f_{a1} \odot m) + f_{a1},
\end{equation}
$\odot$ represents element-wise multiplication. It is used to combine the audio feature \(f_{a1}\) and the generated mask \(m\). This operation effectively suppresses noise components within the audio while preserving valuable speech information. Subsequently, the enhanced audio feature \(\hat{f}_a\) is combined with the original audio feature \(f_{a1}\), resulting in the final enhanced audio information, denoted as \(\text{enhanced}_a\).

\subsection{Audio-aware Visual Refinement Module} \label{subsec:audio-to-video}

Firstly, the features for each modality---audio features \( f_{a2} \in \mathbb{R}^{T_1 \times C_1} \) and visual features \( f_v \in \mathbb{R}^{T_1 \times C_1 \times H_1 \times W_1} \)---are obtained through specific frontend processing. 
Audio features are downsampled, averaged, and replicated to retain semantics and align with the visual modality. Studies show that audio provides both content and spatial cues \cite{gaver1993world}, and audio-visual synchrony aids sound localization \cite{hershey1999audio}. Inspired by this, we propose the Audio-aware Visual Refinement Module (AVRM), which leverages audio to refine visual features, effectively mitigating occlusion and synchronization issues with minimal parameter overhead.
% The audio features undergo convolutional downsampling, followed by temporal averaging and replication, which effectively capture semantic content while maintaining alignment with the visual stream. Psychological and physiological studies have demonstrated that sound not only provides information about its source but also its location \cite{gaver1993world}. Building on this, Hershey et al. \cite{hershey1999audio} further explored how audio-visual synchrony could be used to localize sound sources. Their research indicated that the high correlation between audio and visual modalities facilitates the identification of visual areas closely associated with the audio signals. Inspired by these findings, we developed the Audio-aware Visual Refinement Module (AVRM) for Audio-to-Visual Enhancement, which leverages audio signals to model visual areas.

\begin{figure*}[t!]
    \centering
    \includegraphics[width=0.7\linewidth]{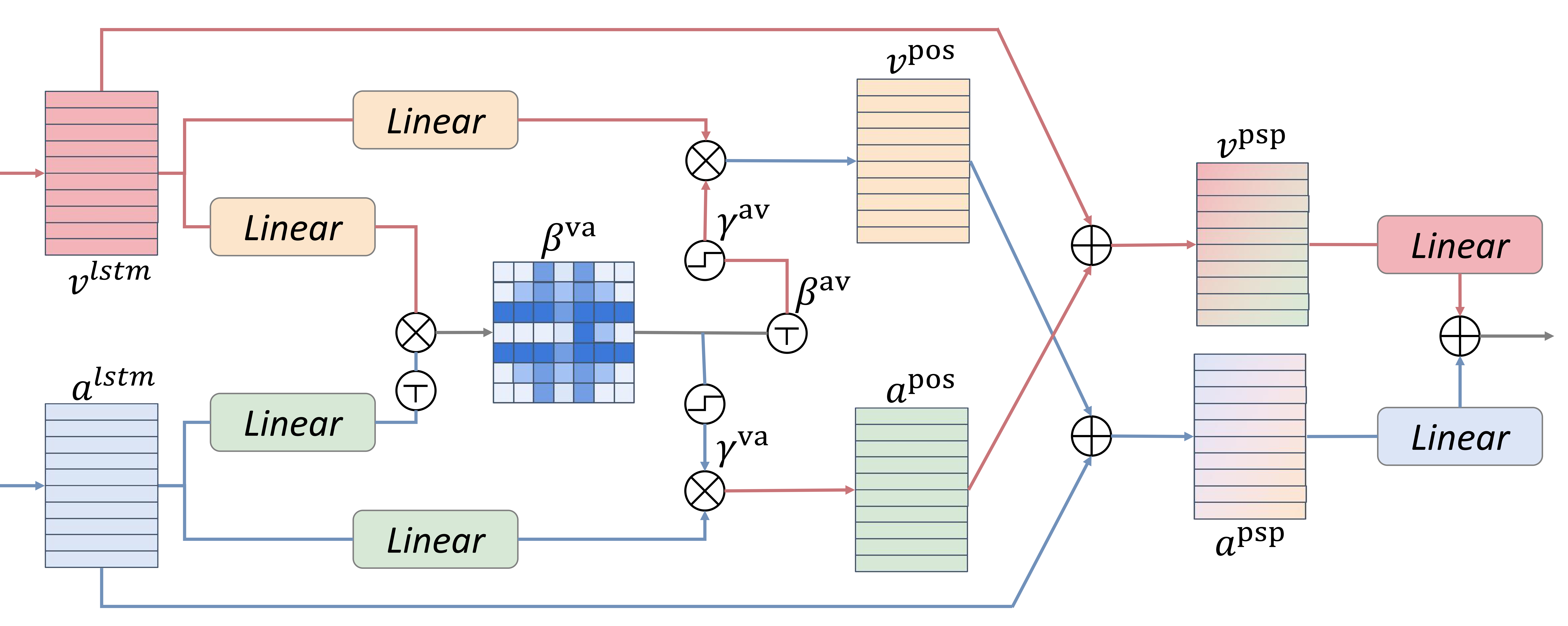}
    \caption{An illustration of the threshold-based selection mechanism.}
    \label{fig:psp-main}
\end{figure*}

Given the exceptional performance of attention mechanisms \cite{vaswani2017attention} in various application domains, such as image captioning, we decided to utilize this mechanism for audio-assisted visual enhancement. Each visual frame is initially divided into \( k \) regions.
% , and the optimal value of \( k \) is determined through experiments shown in Table \ref{tab:k_ablation}. 
In this attention framework, the input audio \( f_{a2} \) and visual features \( f_{v} \) are merged in the following manner:
\begin{equation}
\text{Context}_v = W_3 \sigma \left( W_1 \cdot \text{flatten}_{H,W}(f_v) + (W_2 \cdot f_{a2}) \mathbf{1}^T \right),
\end{equation}
\begin{equation}
w_t = \text{Softmax}(\text{Context}_v),
\end{equation}
\( W_1 \in \mathbb{R}^{k \times C_1} \), \( W_2 \in \mathbb{R}^{k \times C_1} \), and \( W_3 \in \mathbb{R}^{1 \times k} \) serve as linear dimension transformation functions, where \( k \) denotes the number of visual regions. All entries in \( \mathbf{1} \in \mathbb{R}^{k} \) are ones, used to broadcast audio features during the fusion process. Additionally, the operation \( \text{flatten}_{H,W} \) compresses the spatial dimensions of visual information into a single dimension, and \( \sigma(\cdot) \) is the hyperbolic tangent function, employed to process the merged features. The computed attention weights \( w_t \in \mathbb{R}^k \) allow the model to focus on visual regions that are highly correlated with the audio signal locations. \( w_t \) is derived using the softmax activation function. Subsequently, the attention function \( f_{\text{att}} \) is defined, which adaptively learns from the input audio features \( f_{a2} \) and visual features \( f_v \), to produce enhanced visual features \( \text{enhanced}_v \in \mathbb{R}^{T_1 \times D_1} \). 
\begin{equation}
\text{enhanced}_{v_t} = f_{\text{att}}(f_{a2}, f_v) = \sum_{i=1}^{k} w_t^i {(f_v)}_t^i.
\end{equation}

At each time step \( t \), the visual content vector \( \text{enhanced}_{v_t} \) is calculated. \( t \) represents the \( t \)-th time step, and \( i \) indicates the \( i \)-th visual region. This method of audio-guided visual region weighting fusion enhances the model's ability to perceive and focus on semantically relevant areas in complex and visually corrupted scenarios.

\subsection{Threshold-based Selection Mechanism} \label{subsec:threshold}
\begin{figure}[t!]
\centering
\includegraphics[width=1\linewidth]{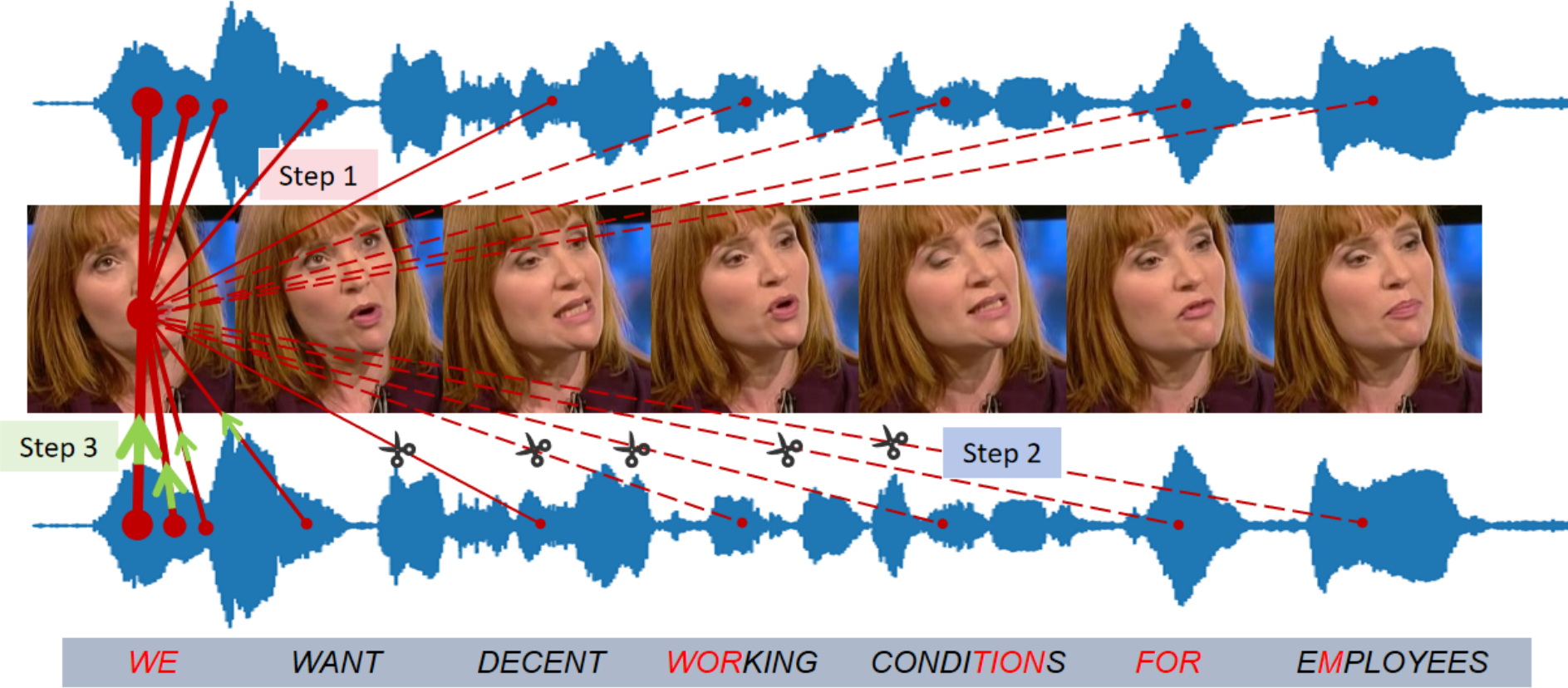} % 替换为你的图片文件名
\caption{This figure illustrates the threshold-based selection mechanism. In this example, only the first segment contains `WE', with three strong audio-related connections shown. Red lines represent audio-visual pairings—solid for relevant pairs and dotted for irrelevant ones. Line thickness indicates pairwise similarity. `WE' shows negative connections with the following words (`WORKING CONDITIONS FOR EMPLOYEES'), positive ones with preceding words (including `WANT'), and a weak link to `DECENT'. `Step 1' outlines the construction of all-pair connections, `Step 2' involves pruning negative and weak connections, and `Step 3' denotes feature aggregation.}
\label{fig:psp}
% \vspace{-0.5em}
\end{figure}

Threshold-based Selection Mechanism (TBSM) enables networks to learn more representative features by leveraging the similarity between audio and visual features.  The TBSM structure is shown in Figure \ref{fig:psp-main}. Inspired by methods described in \cite{zhou2021positive}, TBSM involves three key steps. Figure \ref{fig:psp} visualizes the three steps of the TBSM.

The first step is all-pair connection construction. In AVSR tasks, audio-visual pairs are interconnected on a frame-by-frame basis. As described in the previous section, the enhanced audio features, denoted as \( \text{enhanced}_{a} \), and visual features, denoted as \( \text{enhanced}_{v} \), first undergo preliminary temporal modeling using a bidirectional LSTM layer. This process produces \( a^{\text{lstm}} \in \mathbb{R}^{T_1 \times 2D_1} \) and \( v^{\text{lstm}} \in \mathbb{R}^{T_1 \times 2D_1} \), which are then used to calculate the strength of the fully-connected pairs, detailed in the subsequent formulas.
\begin{equation}
\beta^{va} = \frac{(\mathbf{v}^{\text{lstm}} \mathbf{W}^v_1) (\mathbf{a}^{\text{lstm}} \mathbf{W}^a_1)^T}{\sqrt{2D_1}}, \quad \beta^{av} = (\beta^{va})^T,
\end{equation}
the parameters $W_1^v$ and $W_1^a$ represent learnable parameters of the linear transformations, implemented by corresponding linear layers. Here, $2D_1$ denotes the dimensionality of both visual and audio features, and $T$ stands for the transpose operation. The term $\beta^{va} \in \mathbb{R}^{T_1 \times T_1}$ calculates the connectivity strength of visual features to each audio feature. Conversely, $\beta^{av} \in \mathbb{R}^{T_1 \times T_1}$ computes the connectivity strength of audio features to each visual feature, reflecting the degree of mutual influence between modalities.
% The second step involves pruning negative and weak connections. The connections built in the first step are categorized into positive, weak, and negative types based on their similarity values. Similar to the AVE classification task \cite{zhou2021positive}, the AVSR task is also optimized by establishing similarities between the motion images of lips and corresponding audio sounds. Key lip formations correspond with critical phonemes to establish strong connections. This step aims to collect more positive connections, while weak and negative connections formed by noise and background information in the images are identified and pruned. In this step, we first apply the ReLU activation function to activate all audio-visual pairs, followed by L1 normalization to obtain the normalized similarity matrices $\beta^{va}$ and $\beta^{av}$. We further apply a predefined threshold for feature selection, retaining strongly correlated audio-visual pairs.
The second step prunes weak and negative connections. Based on similarity scores, connections are classified as positive, weak, or negative—similar to the AVE classification in~\cite{zhou2021positive}. In AVSR, strong links are formed between key lip movements and corresponding phonemes. This step aims to retain positive connections while removing those caused by noise or background. We apply ReLU to activate audio-visual pairs, then use L1 normalization to obtain similarity matrices \( \beta^{va} \) and \( \beta^{av} \). Finally, a threshold is used to select strongly correlated pairs.

\begin{equation}
\begin{aligned}
\gamma^{va} &= \beta^{va} \mathcal{I}(\beta^{va} - \tau), \\
\gamma^{av} &= \beta^{av} \mathcal{I}(\beta^{av} - \tau),
\end{aligned}
\end{equation}
$\tau$ is a hyperparameter that controls how many connections are pruned. The indicator function $\mathcal{I}(\cdot)$ returns 1 if the input is $\geq 0$, and 0 otherwise. Audio-visual pairs with similarity scores $\geq \tau$ are kept as strong connections, while others are removed. After thresholding, a second L1 normalization is applied to generate the refined similarity matrices $\gamma^{va}$ and $\gamma^{av} \in \mathbb{R}^{T_1 \times T_1}$.

% $\tau$ serves as a hyperparameter that dictates the quantity of connections to be trimmed. $\mathcal{I}(\cdot)$ functions as an indicator, returning $1$ when the input is equal to or greater than $0$, and $0$ otherwise. Consequently, audio-visual pair connections with similarity scores meeting or exceeding $\tau$ are classified as strong connections; those falling below are deemed weak or negative and are severed. Following the thresholding process, the system executes another round of L1 normalization to produce the refined similarity weight matrices $\gamma^{va}$ and $\gamma^{av}$, with dimensions $R^{T_1 \times T_1}$.

The final step is feature aggregation. After the initial two stages, we have identified audio (visual) frames that are highly correlated with visual (audio) frames. Similar to the roles played by the AVRM and CMNSM, this phase further enhances bidirectional cross-modal complementarity, using visually correlated frames to augment audio information, and vice versa. Utilizing the established similarity weight matrices $\gamma^{va}$ and $\gamma^{av}$, the audio and visual features are updated as follows:
\begin{equation}
\begin{aligned}
a^{\text{psp}} &= \overbrace{\gamma^{\text{av}} \left( v^{\text{lstm}} W_2^v \right)}^{v^{\text{pos}}} + a^{\text{lstm}}, \\
v^{\text{psp}} &= \overbrace{\gamma^{\text{va}} \left( a^{\text{lstm}} W_2^a \right)}^{a^{\text{pos}}}  + v^{\text{lstm}},
\end{aligned}
\end{equation}
$W_2^v \in \mathbb{R}^{2D_1 \times 2D_1}$ and $W_2^a \in \mathbb{R}^{2D_1 \times 2D_1}$ represent learnable parameters of the linear transformations. The $v^{\text{pos}}$ captures visual frames that are highly correlated with the audio signals. These visual features, when combined with audio features, further enhance the audio representation, resulting in the audio feature $a^{\text{psp}} \in \mathbb{R}^{T_1 \times 2D_1}$, and similarly for visual features $v^{\text{psp}}$. This approach enables the extraction of more distinctive features. Subsequently, features from both modalities are processed through individual linear layers before being fused, ultimately yielding the fused feature $fusion \in \mathbb{R}^{T_1 \times 2D_1}$.

% \vspace{-2pt} %
\subsection{Loss Calculation} \label{subsec:loss}
We train our AD-AVSR model end-to-end using CTC \cite{hannun2014deep}  loss, commonly adopted in AVSR. After a threshold-based selection, the fused feature has dimension \( \mathbb{R}^{T \times 2D_1} \), which is reduced via two linear projections to \( f_0 \in \mathbb{R}^{T \times \frac{D_1}{2}} \).
Let \( x = [x_1, \ldots, x_T] \) and \( y = [y_1, \ldots, y_L] \) denote the input and target sequences. We use a 6-layer Conformer~\cite{burchi2023audio}, an enhanced Transformer~\cite{vaswani2017attention} with convolution to capture intra- and inter-modal temporal dependencies. The output \( f_1 \in \mathbb{R}^{T \times \frac{D_1}{2}} \) is used to compute the CTC loss:
\[
p_c(y|x) \approx \prod_{t=1}^{T} p(y_t|x).
\]
CTC assumes conditional independence and enforces monotonic alignment, making it well-suited for lip-reading tasks.

% Ultimately, the sequence feature $f_1$ is passed into a decoding structure composed of six layers of transformer decoders, which delve deeper into the sequence features and transcribe them into corresponding lip reading sentence text sequences, while also computing the attention loss \cite{watanabe2017hybrid} to optimize model performance, defined as \(p_a(y|x) = \prod_{l=1}^L p(y_l | y_{<l}, x)\). This loss mechanism fundamentally computes the KL divergence. Compared to CTC loss, it captures the global context information of the input sequence related to the current output, thereby generating more accurate and coherent results. Similar to previous studies \cite{ma2021end,hong2022visual,hong2023watch}, we combine these two types of losses and calculate them according to a specific ratio. The formula is as follows:
The sequence feature \( f_1 \) is fed into a six-layer Transformer decoder, which further models the sequence and generates the corresponding lip-reading text. During this process, we compute the attention loss~\cite{watanabe2017hybrid}, defined as
\[
p_a(y|x) = \prod_{l=1}^L p(y_l | y_{<l}, x).
\]

This loss captures global context via KL divergence, producing more accurate and coherent outputs than CTC. Following prior works~\cite{ma2021end,hong2022visual,hong2023watch}, we combine CTC and attention losses using a weighted sum, formulated as:
\begin{equation}
\mathcal{L} = \lambda \log p_a(y|x) + (1-\lambda) \log p_c(y|x),
\end{equation}
\(\lambda\) is the weight parameter balancing the two losses.

\section{Experiments}
\label{sec:expers}
\subsection{Implementation Details} \label{sec:implementation}

\begin{table*}[ht]
\centering
\caption{Comparison with state-of-the-art methods on the LRS2 and LRS3 under different SNR conditions. `Clean' means the test set contains no added noise. `10', `5', `0', and `-5' denote SNR levels applied to the test set. `Avg' is the average WER (\%) across these five conditions.}
\label{tab:main}
\begin{tabular}{c|ccccccc|cccccc}
\toprule
 \multirow{2}{*}{\textbf{Input Model}} & \multirow{2}{*}{\textbf{Method}} & \multicolumn{6}{c|}{\textbf{LRS2}} & \multicolumn{6}{c}{\textbf{LRS3}} \\
 & & clean & 10 & 5 & 0 & -5 & avg & clean & 10 & 5 & 0 & -5 & avg \\
\midrule
A & ASR~\cite{ma2021end} & 4.9 & 7.4 & 8.7 & 10.2 & 29.1 & 12.1 & -- & -- & -- & -- & -- & -- \\
A & AVEC~\cite{burchi2023audio} & 3.1 & 7.6 & 8.6 & 27.0 & 70.5 & 19.5 & 2.3 & 4.1 & 9.3 & 32.4 & 75.9 & 20.7 \\
\cdashline{1-14}
% V & VSR~\cite{ma2021end} & 4.6 & 7.8 & 10.8 & 16.5 & 24.9 & 12.9 & 3.2 & 5.4 & 8.3 & 14.6 & 22.3 & 10.8 \\
A + V & Conformer~\cite{ma2021end} & 4.6 & 7.8 & 10.8 & 16.5 & 24.9 & 12.9 & 3.2 & 5.4 & 8.3 & 14.6 & 22.3 & 10.8 \\
A + V & V-CAFE~\cite{hong2022visual} & 4.3 & 5.5 & 6.4 & 11.0 & 22.4 & 9.9 & 2.9 & 4.0 & 8.4 & 12.5 & 19.3 & 9.4 \\
A + V & AV-Hubert~\cite{shi2022robust} & --  & --  & --  & --   & --   & -- & 2.0 & 2.1 & \textbf{2.6} & 5.8 & 16.6 & 5.8 \\
A + V & AVEC~\cite{burchi2023audio} & 2.6 & 2.8 & \textbf{3.4} & 5.0 & 9.7  & 4.7 & 2.0 & 2.5 & 3.1 & 4.9 & 11.2 & 4.7 \\
A + V & AV-Relscore~\cite{hong2023watch} & 4.1 & 4.3 & 5.2 & 6.2 & 11.3 & 6.2 & 2.8 & 2.9 & 3.2 & 4.8 & 8.7 & 4.5 \\
A + V & A+VH~\cite{zhang2024visual} & 2.6 & 3.1 & 3.9 & 7.1 & 12.6 & 5.9 & 2.2 & 2.2 & 3.4 & 6.4 & 14.3 & 5.7 \\
\midrule
A + V & \textbf{AD-AVSR (ours)} & \textbf{2.4} & \textbf{2.8} & 3.6 & \textbf{6.0} & \textbf{9.4} & \textbf{4.8} & \textbf{2.0} & \textbf{2.1} & 3.2 & \textbf{4.7} & \textbf{8.2} & \textbf{4.0} \\
\bottomrule
\end{tabular}
\end{table*}
\begin{table*}[t]
\centering
\caption{WER (\%) comparisons with state-of-the-art methods under audio-visual corrupted environments on LRS2. The first row represents visual corruption types (patch occlusion and noise), and the second row indicates audio noise levels (SNR in dB).}
\label{tab:r2}
\begin{tabular}{lcccccccccccc}
\hline
\multirow{2}{*}{\textbf{Method}} & \multicolumn{6}{c}{\textbf{Occlusion}} & \multicolumn{6}{c}{\textbf{Noise}} \\
& clean & 10 & 5 & 0 & -5 & avg & clean & 10 & 5 & 0 & -5 & avg \\
\hline
Conformer~\cite{ma2021end}     & 4.9 & 8.0 & 10.8 & 16.6 & 25.1 & 13.1 & 4.8 & 7.8 & 10.7 & 16.8 & 25.7 & 13.2 \\
V-CAFE~\cite{hong2022visual}   & 4.4 & 5.7 & 6.5  & 11.3 & 22.4 & 10.0 & 4.9 & 5.6 & 6.4  & 11.4 & 22.8 & 10.2 \\
AV-RelScore~\cite{hong2023watch} & 4.2 & 4.4 & 5.2  & 6.4  & 11.3 & 6.3  & 4.5 & 4.3 & 5.1  & 6.4  & 11.2 & 6.3 \\
\textbf{AD-AVSR (ours)}        & \textbf{2.5} & \textbf{3.0} & \textbf{3.7} & \textbf{6.2} & \textbf{9.5} & \textbf{5.0} & \textbf{2.8} & \textbf{3.2} & \textbf{3.7} & \textbf{6.3} & \textbf{9.5} & \textbf{5.1} \\
\hline
\end{tabular}
\end{table*}

% \begin{table*}[ht]
% \centering
% \caption{Comparison with state-of-the-art methods on LRS2 and LRS3 under different SNR conditions. Lower is better.}
% \label{tab:main}
% \begin{tabular}{ccccccc|cccccc}
% \toprule
%  \multirow{2}{*}{Method} & \multicolumn{6}{c|}{LRS2} & \multicolumn{6}{c}{LRS3} \\
%  & clean & 10 & 5 & 0 & -5 & avg & clean & 10 & 5 & 0 & -5 & avg \\
% \midrule
% Conformer~\cite{ma2021end} & 4.6 & 7.8 & 10.8 & 16.5 & 24.9 & 12.9 & 3.2 & 5.4 & 8.3 & 14.6 & 22.3 & 10.8 \\
% V-CAFE~\cite{hong2022visual}     & 4.3 & 5.5 & 6.4 & 11.0 & 22.4 & 9.9 & 2.9 & 4.0 & 8.4 & 12.5 & 19.3 & 9.4 \\
% AV-Hubert~\cite{shi2022robust}  & --  & --  & --  & --   & --   & -- & 2.0 & 2.1 & \textbf{2.6} & 5.8 & 16.6 & 5.8 \\
% % P\&U Net~\cite{}    & 3.6 & 4.0 & 5.8 & 11.2 & 27.8 & 10.5 & 3.0 & --  & --  & --   & --   & -- \\
% AVEC~\cite{burchi2023audio}      & 2.6 & 2.8 & \textbf{3.4} & 5.0 & 9.7  & 4.7 & 2.0 & 2.5 & 3.1 & 4.9 & 11.2 & 4.7 \\
% AV-Relscore~\cite{hong2023watch} & 4.1 & 4.3 & 5.2 & 6.2 & 11.3 & 6.2 & 2.8 & 2.9 & 3.2 & 4.8 & 8.7 & 4.5 \\
% A+VH~\cite{zhang2024visual}      & 2.6 & 3.1 & 3.9 & 7.1 & 12.6 & 5.9 & 2.2 & 2.2 & 3.4 & 6.4 & 14.3 & 5.7 \\
% \midrule
% \textbf{ours}          & \textbf{2.4} & \textbf{2.8} & 3.6 & \textbf{6.0} & \textbf{9.4} & \textbf{4.8} & \textbf{2.0} & \textbf{2.1} & 3.2 & \textbf{4.7} & \textbf{8.2} & \textbf{4.0} \\
% \bottomrule
% \end{tabular}
% \end{table*}
\textbf{Dataset.}
% Our method is evaluated on two widely used AVSR datasets: LRS2~\cite{son2017lip} and LRS3~\cite{afouras2018lrs3}. 
% LRS2 is an English sentence-level dataset with a total duration of approximately 224 hours, consisting of 144{,}482 utterances extracted from 482 BBC video clips. It provides around 1{,}000 clips for validation and about 1{,}200 clips for testing. 
% LRS3 is a larger English sentence dataset with a total duration of about 439 hours. It contains approximately 120{,}000 pre-training samples (408 hours), 32{,}000 samples for training and validation (30 hours), and 1{,}321 samples for testing (0.9 hours).
Our method is evaluated on two widely used AVSR datasets: LRS2~\cite{son2017lip} and LRS3~\cite{afouras2018lrs3}.
LRS2 contains 144,482 English utterances (224 hours) extracted from 482 BBC videos, with around 1,000 clips for validation and 1,200 for testing.
LRS3 is a larger dataset with a total of 439 hours, including about 120,000 pre-training samples (408 hours), 32,000 for training/validation (30 hours), and 1,321 for testing (0.9 hours).

\noindent
\textbf{Architecture Details.} 
We follow the visual frontend from~\cite{hong2023watch}, sampling visual features at 25Hz and audio at 16,000Hz. To fit the AVRM design, we remove global average pooling and retain spatial dimensions. For audio, we apply a dual-stream encoding strategy, both derived from the same waveform.
In the audio-enhanced visual branch, the waveform is transformed into a mel-spectrogram via STFT and processed by a 1D convolution. We then average every 25 frames and replicate the result to enrich each audio frame. In the visual-enhanced audio branch, we use the audio frontend from~\cite{hong2023watch}, allowing each visual frame to guide audio enhancement effectively.
The bidirectional modality enhancement module is our main innovation. In AVRM, the visual region parameter \( k \) is set to 9. After AVRM and CMNSM, a bidirectional LSTM layer performs initial temporal modeling. The audio and visual features are then refined by the TBSM module, with threshold \( \tau = 0.095 \) as in~\cite{zhou2021positive}.

Audio and visual features are fused into a 1024-dimensional representation, followed by a 6-layer Conformer encoder for CTC loss computation. The output is then passed to a 6-layer Transformer decoder (4 attention heads, hidden size 256) to generate predictions, where an attention-based loss is also computed. The final loss is a weighted sum of the CTC and attention losses, with \( \lambda = 0.9 \).

% The audio and visual representations are then fused along the feature dimension, resulting in a combined feature of size 1024. A 6-layer Conformer encoder is employed on the fused features, and the CTC loss is calculated based on its output. Subsequently, the encoder output is fed into a 6-layer Transformer decoder, which consists of 4 attention heads and a hidden dimension of 256, to generate the final predictions. An attention-based loss is computed at this stage. The CTC and attention-based losses are then combined using a fixed weight ($\lambda = 0.9$) to compute the final loss. 
% The hybrid loss jointly captures the temporal structure of lip reading and semantic information, leveraging their complementary strengths.

\noindent
\textbf{Preprocessing.} 
In preprocessing, we use dlib to detect 68 facial landmarks and apply interpolation for missing frames. Faces are aligned via affine transformation to reduce rotation and scale variations. Visual data is augmented with random cropping ($112 \times 112$), and later resized to $96 \times 96$ with horizontal flipping (p = 0.5). Pixels are normalized to $[0, 1]$ and standardized (mean = 0.421, std = 0.165).
For audio, random noise with SNR $\in {-5, 0, 5, 10, 15}$ is added during training, and waveforms are normalized. We further apply SpecAugment (Time Masking) and Cutout to introduce temporal and spatial perturbations. All augmentations are disabled during evaluation, where only normalization is applied.
% In the preprocessing stage, to enhance both video and raw waveform data, we use dlib to detect and track 68 facial landmarks. For frames where landmarks are not detected, interpolation is applied. To eliminate variations caused by facial rotation and scale, an affine transformation is used to align faces to a canonical coordinate space.
% To augment the visual data, random cropping of size $112 \times 112$ is applied to each video frame. During training, random noise is added to the audio waveform with the signal-to-noise ratio (SNR) randomly selected from ${-5, 0, 5, 10, 15}$. Each raw waveform is normalized by subtracting its mean and dividing by its standard deviation. For the visual stream, pixel values of input frames are normalized to the $[0, 1]$ range, followed by random cropping to a fixed resolution of $96 \times 96$ and random horizontal flipping with a probability of 0.5 to increase spatial diversity. Dataset-specific normalization is then applied using a mean of 0.421 and a standard deviation of 0.165.
% To introduce temporal and spatial perturbations, we further apply Time Masking \cite{park2019specaugment} and Cutout during training, which randomly mask regions along the temporal axis and spatial area, respectively. These augmentation operations are disabled during evaluation to ensure consistency. In the validation and testing phases, audio preprocessing is limited to the aforementioned normalization step.

\noindent
\textbf{Training.} We adopt a curriculum learning strategy~\cite{zhou2021positive}. 
% The model is first trained using video clips with lengths less than 100 frames. Subsequently, training is performed with samples of up to 150, 300, and 600 frames in length. After four stages of training, the final model is obtained. The number of training epochs for each stage is set to 50, 50, 20, and 20, respectively.
We use the Adam optimizer \cite{kingma2014adam} with the Noam learning rate scheduler, which includes a warm-up phase and scales the learning rate based on the model dimension.  All experiments are conducted on four NVIDIA RTX 3090 GPUs, with each experiment on the LRS2 dataset taking approximately eight days to train.

% \subsection{Main Result} \label{sec:outcome}

\subsection{Performance Comparison and Noise Robustness Evaluation} \label{subsec:comparision}

\noindent 
We first compare the proposed AVSR audio-visual enhancement framework with state-of-the-art methods, including AV-Hubert \cite{shi2022robust}, AV-RelScore \cite{hong2023watch} and A+VH \cite{zhang2024visual}, on the LRS2 and LRS3 datasets. To evaluate model robustness, we further conduct audio corruption modeling on both datasets, where the signal-to-noise ratio (SNR) ranges from -5 to 10 dB, along with a clean audio condition. The comparison results on both datasets are presented in Table~\ref{tab:main}. Under clean audio conditions, our method achieves the best performance on both datasets. Moreover, in scenarios with varying levels of audio noise, the proposed method consistently performs well in most cases, demonstrating strong robustness to audio interference.

In addition to adding different levels of SNR, we also followed the settings in \cite{hong2023watch} to introduce occlusion and noise to the visual frames. Moreover, we conducted additional robustness experiments on the LRS2 dataset with simultaneous audio-visual corruption. As shown in Table~\ref{tab:r2}, our proposed AD-AVSR still demonstrates strong robustness.

\begin{table*}[t]
\centering
\caption{Parameters, FLOPs, Total Training Time, Word Error Rate (WER), and Character Error Rate (CER) of AD-AVSR on the LRS2 and baseline at SNR = -5. '→' denotes sequential addition, and values in parentheses indicate the relative increase or decrease compared to the baseline.}
\label{tab:r1}
\begin{tabular}{lccccc}
\hline
\textbf{Method} & \textbf{Params (10$^6$)} & \textbf{FLOPs (10$^9$)} & \textbf{Total Hours (h)} & \textbf{WER (\%)} & \textbf{CER (\%)} \\
\hline
baseline            & 74.05           & 86.86           & 192.0 & 24.9 & 13.6 \\
$\rightarrow$ +AVRM & 74.59 (+0.7\%)  & 87.48 (+0.7\%)  & 193.4 (+0.7\%) & 20.2 (↓18.9\%) & 12.4 (↓8.8\%) \\
$\rightarrow$ +CMNSM & 76.42 (+3.2\%) & 87.85 (+1.1\%)  & 194.2 (+1.1\%) & 19.3 (↓22.5\%) & 9.5 (↓30.1\%) \\
$\rightarrow$ \textbf{+TBSM (AD-AVSR)} & 90.07 (+31.6\%) & 88.83 (+2.3\%)  & 196.4 (+2.3\%) & \textbf{9.4} (↓62.2\%) & \textbf{4.9} (↓64.0\%) \\
\hline
\end{tabular}
\end{table*}
\begin{table}[t!]
\centering
\caption{Ablation study on the effectiveness of AVRM, CMNSM, and TBSM modules on the LRS2. WER and CER stand for Word Error Rate and Character Error Rate, respectively.}
\label{tab:ablation}
\begin{tabular}{ccccccr}
\toprule
\textbf{baseline} & \textbf{AVRM} & \textbf{CMNSM} & \textbf{TBSM} & \textbf{WER (\%)} & \textbf{CER (\%)} \\
\midrule
\ding{51} & \ding{55} & \ding{55} & \ding{55} & 24.9 & 13.6 \\
\ding{51} & \ding{51} & \ding{55} & \ding{55} & 20.2 & 12.4 \\
\ding{51} & \ding{55} & \ding{51} & \ding{55} & 22.4 & 12.8 \\
\ding{51} & \ding{55} & \ding{55} & \ding{51} & 16.3 & 8.0 \\
\ding{51} & \ding{51} & \ding{51} & \ding{55} & 19.3 & 9.5 \\
% \ding{51} & \ding{51} & \ding{55} & \ding{51} & 14.2 & -- \\
% \ding{51} & \ding{55} & \ding{51} & \ding{51} & 14.9 & -- \\
\ding{51} & \ding{51} & \ding{51} & \ding{51} & \textbf{9.4} & \textbf{4.9} \\
\bottomrule
\end{tabular}
\end{table}
\begin{figure}[t!]
\centering
\includegraphics[width=0.8\linewidth]{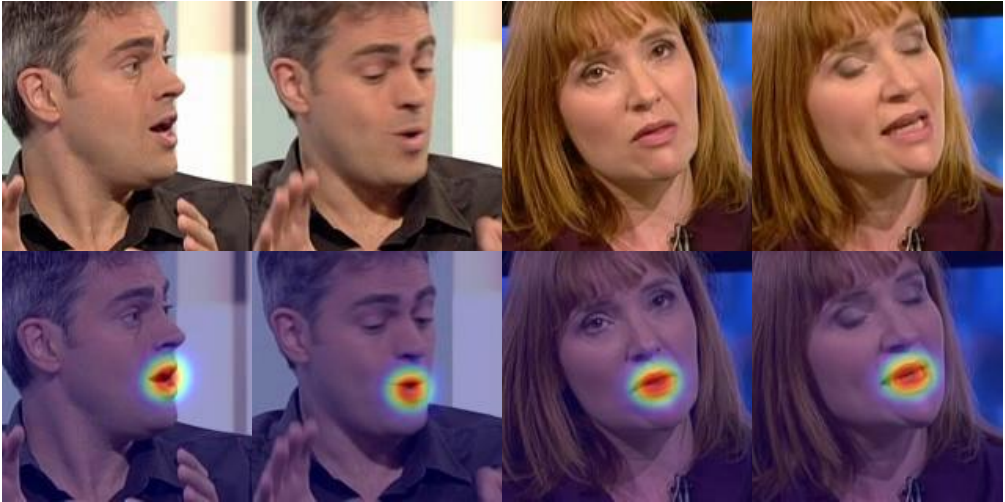} % 替换为你的图片文件名
\caption{Qualitative visualization of the Audio-aware Visual Refinement Module, which adaptively focuses on the lip region based on audio articulation to enhance detection of lip movement variations.}
\label{fig:heat}
\vspace{-1.0em}
\end{figure}

\subsection{Ablation Study}
\label{subsec:abalation}
We conducted an ablation study on LRS2 under noisy conditions (SNR = -5 dB) to evaluate different model variants.

% Starting from a baseline network without any enhancement components, we incrementally added CMNSM, AVRM, TBSM, and their combination. The word error rate (WER) improved from 24.9\% to 9.4\%, as summarized in Table~\ref{tab:ablation}. The results demonstrate that compared to the model without any audio or visual enhancement, the bidirectional modality enhancement module yields consistent improvements, indicating the benefit of leveraging complementary information from other modalities. Overall, the proposed method achieves a 15.5\% reduction in word error rate compared to the advanced baseline methods.

Starting from a baseline network without any enhancement components, we incrementally added CMNSM, AVRM, TBSM, and their combination, leading to a consistent reduction in word error rate (WER) from 24.9\% to 9.4\%, as shown in Table~\ref{tab:ablation}. While our proposed AD-AVSR model increases the number of parameters from 74.05M to 90.07M, it maintains comparable FLOPs (+2.3\%) and training time in Table~\ref{tab:r1}. Despite the moderate increase in model complexity, AD-AVSR achieves substantial performance gains under severe noise conditions (SNR = -5), reducing WER by 62.2\% (from 24.9\% to 9.4\%) and character error rate (CER) by 64.0\% (from 13.6\% to 4.9\%). These results highlight that the added complexity is well-justified by the significant improvements, demonstrating the effectiveness of the proposed bidirectional modality enhancement in leveraging complementary cross-modal information.

For the audio-to-visual modeling branch, the AVRM module also adopts a cross-modal attention mechanism. It divides each video frame into $k$ spatial regions and adaptively learns which regions correspond to the relevant sound activities. The qualitative visualization of the module is presented in Figure \ref{fig:heat}, with an increased focus on the primary articulatory part (the lips) in the visual representation. According to the results in Table~\ref{tab:ablation}, AVRM achieves a lower word error rate compared to using CMNSM. Audio often contains finer details than visuals, so using it to enhance visual features improves performance. AVRM preserves the spatial information from the visual frontend, and combining spatial and temporal cues makes audio-enhanced visual modeling more effective.

% In the proposed network, the hyperparameter $k$ controls the number of visual regions each video frame is divided into. During training, the network learns to assign higher attention weights to regions that are more relevant to the sound activity. The influence of $k$ on system accuracy under clean audio conditions is shown in Table~\ref{tab:k_ablation}. We observe that increasing the value of $k$ initially leads to consistent performance improvements. However, when $k$ reaches 25, no further gains are observed. Considering the increased computational cost associated with $k=25$, we set $k=16$ as the default number of visual regions for each frame.
% \begin{table}[th]
% \centering
% \caption{Effect of the number of visual regions $k$ on performance under clean audio conditions on LRS2.}
% \label{tab:k_ablation}
% \begin{tabular}{lcccc}
% \toprule
% $k$ & 4 & 9 & 16 & 25 \\
% \midrule
% WER & 5.32 & 3.07 & \textbf{2.41} & \textbf{2.41} \\
% CER & 2.72 & 1.55& \textbf{1.43} & 1.46\\
% \bottomrule
% \end{tabular}
% \end{table}
\begin{table}[th]
\centering
\vspace{-5pt} 
\caption{Comparison of different audio encoding methods on LRS2. A1: time-domain encoding (1D-CNN + ResNet18); 
A2: frequency-domain encoding (mel-spectrogram + 1D-CNN with frame averaging and repeating); 
A3: our proposed dual-stream encoding combining A1 and A2.}
\label{tab:audio_encoding}
\begin{tabular}{cccc}
\toprule
Audio encoding method & A1 & A2 & A3 \\
\midrule
\centering WER & 3.63 & 4.04 & \textbf{2.41} \\
\centering CER & 1.82 & 2.36& \textbf{1.43} \\
\bottomrule
\end{tabular}
\end{table}
\subsection{Effects of Dual-Stream Audio Encoding}
\label{subsec:dual-stream}

After processing through their respective frontends, the two modalities are aligned with the same number of frames.

However, simple alignment strategy is not compatible with our bidirectional enhancement design. In practice, visual and audio modalities are frequently misaligned. To address this, the enhancing modality is designed to carry more information per frame, enabling more effective compensation for the other modality. Accordingly, we experimented with two audio encoding strategies, as shown in Table~\ref{tab:audio_encoding}. A3 achieves the best, showing that using a richer enhancing modality (audio or visual) and a lighter enhanced modality (visual or audio) improves fusion. 
A1 is a time-domain encoding scheme that processes audio waveforms using a 1D-CNN followed by ResNet18. A2 is a frequency-domain scheme that converts audio into mel-spectrograms via STFT and applies a 1D convolution.  To ensure each audio frame carries more information than its visual counterpart, audio features are averaged over 25 frames and repeated 25 times. A1 and A2 are each used in both CMNSM and AVRM during ablation, but in A3, they are used separately for different modules. A3 is our proposed dual-stream encoding scheme that combines A1 and A2, where A1 feeds CMNSM (visual-enhanced audio branch) and A2 feeds AVRM (audio-enhanced visual branch).
% A1 is a time-domain encoding scheme, where audio waveforms are processed by a 1D-CNN followed by ResNet18. This is used in the visual-to-audio enhancement path (CMNSM). A2 is a frequency-domain scheme used in the audio-to-visual path (AVRM), where audio waveforms are converted into mel-spectrograms via STFT and passed through a 1D convolution layer. To ensure each audio frame carries more information than its visual counterpart, features are averaged every 25 frames and repeated 25 times. A3 is our proposed audio dual-stream encoding, combining the strengths of A1 and A2.

A1’s time-domain features provide fine-grained visual dynamics per frame, helping to compensate for short-term gaps in the audio signal. In contrast, A2’s frequency-domain features offer structured semantic cues, helping the visual modality focus on speech-related regions. This asymmetric enhancement strategy overcomes the limits of symmetric fusion, allowing each frame to extract richer information from the complementary modality and improving overall cross-modal modeling.

\section{Conclusion and Discussions}
\label{sec:conclusion}

This paper presents AD-AVSR, which significantly boosts AVSR performance by fully leveraging the complementary strengths of visual and audio modalities. It introduces a dual-stream audio encoding strategy to optimize cross-modal enhancement. We further propose a bidirectional enhancement module—AVRM and CMNSM—to better capture lip movements and suppress audio noise. To address irrelevant audio-visual pairings, we introduce a threshold-based selection mechanism that significantly improves fusion efficiency. Extensive comparative experiments, ablation studies, and visualizations validate the effectiveness of AD-AVSR.

This work considers audio-visual asynchrony and examines the impact of audio noise on our model. To address audio-visual misalignment, we introduce a dual-stream audio encoding strategy during the encoding stage, providing richer information to the enhancing modality. While straightforward, this approach proves effective, and future work can explore alternative solutions. We also aim to focus more on various modality degradation scenarios to further enhance recognition performance and model robustness.

% \section*{Acknowledgments}

%-------------------------------%
% 参考文献
%-------------------------------%
\clearpage
\newpage
\bibliographystyle{ACM-Reference-Format}
\bibliography{reference}

\end{document}